\input{aipcheck}

\documentclass[
    ,final            
  ]
  {aipproc}

\layoutstyle{8x11double}

\begin{document}

\title{Collective Effect Studies of a Beta Beam Decay Ring}

\classification{}
\keywords      {Neutrino Oscillation, Beta Beams, Decay Ring, Collective Effects, Impedance}

\author{C. Hansen and G. Rumolo}{
  address={CERN, Geneva, Switzerland}
}




\vspace{-0.2cm}
\begin{abstract}
The Beta Beam, the concept of generating a pure and intense (anti) neutrino beam by letting accelerated radioactive ions beta decay 
in a storage ring called the Decay Ring (DR), is the basis of one of the proposed next generation neutrino oscillation facilities, 
necessary for a complete study of the neutrino oscillation parameter space.
Sensitivities of the unknown neutrino oscillation parameters depend on the DR's ion intensity 
and of its duty factor (the filled ratio of the ring).
Different methods, including analytical calculations and multiparticle tracking simulations, were used to estimate the 
DR's potential to contain enough ions in as small a part of the ring as needed for the sensitivities. 
Studies of transverse blow up of the beams due to resonance wake fields show that a very challenging upper limit of the 
transverse broadband impedance is required to avoid instabilities and beam loss.  
\end{abstract}
\vspace{-0.2cm}
\maketitle

\vspace{-0.2cm}
\section{Introduction}
\label{sec:intro}
\vspace{-0.2cm}

The discovery of neutrino oscillations \cite{SK-OSCILLATION} has confirmed that neutrinos are massive and that
their flavor ($\nu_e$, $\nu_{\mu}$, $\nu_{\tau}$) and mass eigenstates ($\nu_1$, $\nu_2$, $\nu_3$) are mixed.
Neutrino physics is now in an era of precision measurements of the 
parameters that govern these oscillations:
two $\Delta m^2_{ij} \equiv m_{\nu_i}^2 - m_{\nu_j}^2$ parameters 
($|\Delta m^2_{32}|$, $|\Delta m^2_{21}|$), three mixing angles ($\theta_{23}$, $\theta_{12}$, $\theta_{13}$) 
and a CP violating phase ($\delta_{CP}$). 
The most precise determinations to date 
are  $\Delta m_{21}^2 = \left(7.586_{-0.203}^{+0.212}\right)\cdot10^{-5}$~eV$^2$, $\theta_{12} = \left(34.1^{+15.4}_{-14.5}\right) ^{\circ}$ 
\cite{LATEST-SOLAR},
$| \Delta m_{32}^2| = \left(2.43\pm0.13\right) \cdot 10^{-3}$~eV$^2$ 
\cite{MINOS-ATMO} 
and  $\theta_{23} = \left(45\pm3.4\right)^{\circ}$ 
\cite{SK-2009}. 
\begin{figure}[ht]
\includegraphics[angle=0, width= 79.5mm, height= 40mm]{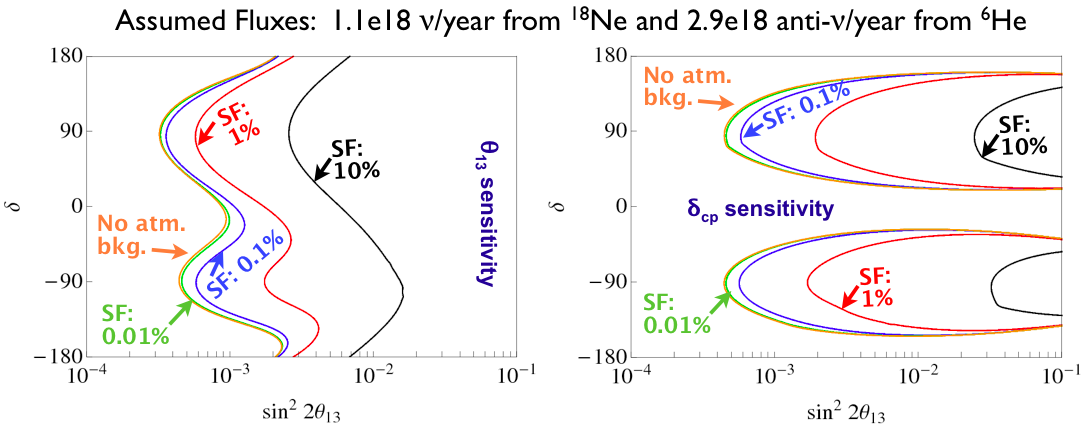} 
\caption{The $\theta_{13}$ (left) and  $\delta_{CP}$ (right)
         sensitivities for different suppression factors (SF)~\cite{FernandezMartinez:2009hb}.  
         A SF of 0.01\% (green) gives almost the same result as the case of no atmospheric background (orange). The 0.1\% SF
         (blue) gives similar sensitivity and with 1\% SF (red) the sensitivity has decreased slightly. }
\label{fig:sens}
\end{figure}
This leaves three unknowns: $\theta_{13}$ (<$12.2^{\circ}$ \cite{CHOOZ-LAST}),
$\delta_{CP}$ and sign($\Delta m_{32}^2$). The discovery of $\theta_{13} > 0$ and $\delta_{CP} \neq 0^{\circ}$ and 
$\neq 180^{\circ}$ would mean existence of CP violation in the leptonic sector. 
Near future experiments will perform precision measurements of the knowns and continue to probe the unknowns
(e.g. T2K started taking data 2009 \cite{Ichikawa:2010zz}).
However, a complete study of all neutrino parameters requires a better characterized neutrino beam with higher flux
then ever available before. 
One of the three present options for a next generation neutrino oscillation facility \cite{FP7-EURONU} 
is the Beta Beam concept \cite{ZUCCHELLI-BETABEAM} wherein it is proposed to store
high energy ($\gamma$~=~100) radioactive ions in a horse-track shaped storage ring, called ``Decay Ring'' (DR), 
with a straight section pointing to a neutrino detector. 
Ions that beta-decay in the straight section emit electron (anti) neutrinos in a pure 
$\nu_e$ ($\bar{\nu}_e$) beam with opening angle $1/\gamma$.
The aimed annual (anti) neutrino fluxes of (1.1e18) 2.9e18~\cite{FP6-FINAL} from ($\beta^-$) $\beta^+$ decaying ($^6$He) $^{18}$Ne ions 
give  $\theta_{13}$ and $\delta_{CP}$ sensitivities shown in fig.~\ref{fig:sens}~\cite{FernandezMartinez:2009hb} for 
different suppression factors (SF) of the detector (which coincides with the duty factor of the DR). 
Fig.~\ref{fig:sens} shows that with the nominal fluxes the beam can only fill less than 1\%  of the DR. 
The challenges of how to produce enough ions, how to accelerate them through a Beta Beam complex and
how to achieve 0.58\% SF are described elsewhere~\cite{FP6-FINAL}. In the studied scenario the Beta Beam complex is 
based at CERN, making use of already existing PS and SPS machines and the DR has the same circumference as SPS, $C$~=~6911.6~m. 
Assuming 20 bunches of 6911.6m$\times$0.58\%/20~=~2~m each, the number of $^{18}$Ne ($^6$He) per bunch have to be
$3\cdot10^{12}$ ($4\cdot10^{12}$) to reach the nominal (anti) neutrino fluxes.
In this report we 
focus on the question whether the required amount of ions can populate such short bunches 
without too large a risk of beam instabilities.
This collective effect study is based on the previous design studies~\cite{FP6-FINAL} and all parameters
used are listed in table~\ref{tab:drValuesIonDep} and \ref{tab:drValues}.

\begin{table}[htbp]
\begin{tabular}{l|cc}
\hline
Parameters & DR $^{18}$Ne & DR $^{6}$He \\ 
\hline
$Z$  & 10  & 2  \\ 
$A$ &  18  & 6  \\ 
$V_{RF}$ [MV] & 11.96 & 20.00 \\ 
$E_{rest}$ [MeV] &  16767.10  & 5605.54  \\ 
$N_{B}$ &  3.1e+12  & 4.0e+12  \\ 
$t_{1/2}$ [s] &  1.67  &   0.81  \\ 
$T_{c}$ [s] &   3.60  &   6.00  \\ 
\hline
$r_{0}$ [m] $= r_{p}Z^{2}/A$ &  8.53e-18  & 1.02e-18  \\ 
$E_{tot}$ [GeV] $= \gamma \cdot E_{rest}$ &  1676.71  & 560.55  \\ 
$\hat{I}$ [A] $= ZeN_{B}/ \tau_{b}$ & 755.80  & 195.04  \\ 
$I_{b}$ [A] $= ZeN_B  /T_{rev}$ &   0.22  &   0.06  \\ 
$\varepsilon_{l}^{^{2\sigma}}$ [eVs] $=\frac{\pi}{2}\beta^{2}E_{tot}\tau_{b}\delta_{max}$ &  43.27  &  14.46  \\ 
\hline
\end{tabular}
\caption{Input parameters (some from~\cite{FP6-FINAL}) above the line
  and calculated parameters below the line. 
}
\label{tab:drValuesIonDep}
\end{table}

\begin{table}[htbp]
\begin{tabular}{l|lc}
\hline
Parameters & {\it Description} & {\it Value}\\ 
\hline
$h$ & {\it Harmonic Number} & 924 \\ 
$C$ [m] & {\it Circumference} & 6911.6 \\ 
$\rho$ [m] & {\it Magnetic Radius} &  155.6 \\ 
$\gamma_{tr}$ & {\it Gamma at Transition} & 27.00 \\ 
$\gamma$ & {\it Relativistic Gamma} &  100.0  \\ 
$\delta_{max}$ & {\it Max. Mom. Spread} & 2.5e-3  \\ 
$L_{b}$ [m] & {\it Full Bunch Length} &  1.970  \\ 
$Q_{y}$ & {\it Vertical Tune} &  12.16  \\ 
$\langle \beta \rangle_{y}$ [m] & {\it Av. Ver. $\beta$tron Func.} & 173.64  \\ 
$b_{y}$ [cm] & {\it Ver. Beam Pipe Size} &    16.0  \\ 
\hline
$Q_{\perp}$ & {\it Quality Factor} &    1.00  \\ 
$\omega_{r,\perp}$  [GHz] & {\it Ang. Resonance Freq.}  &   6.28  \\ 
$R_{s,\perp}  [M \Omega /m]$ & {\it Shunt Impedance}  &  2.00  \\ 
\hline
$\eta = \gamma_{tr}^{-2} - \gamma^{-2}$ & {\it Phase Slip Factor} & 1.27e-3  \\ 
$T_{rev}$ [$\mu$ s] $= C/(\beta c)$ & {\it Revolution Time}  & 23.0558  \\ 
$\omega_{rev}$ [MHz] $= \frac{2 \pi}{T_{rev}}$ & {\it Ang. Revolution Freq.} &   0.27  \\ 
$Q_{s}=\sqrt{\frac{hZeV|\eta \cos\phi_{s}|}{2\pi\beta^{2}E_{tot}}}$ & {\it Synchrotron Tune}  &   3.63e-3  \\ 
$\omega_{c}$ [GHz] $= \frac{\beta c}{b_{x,y}}$ & {\it Cut-Off Ang. Frequency} &    1.87  \\ 
\hline
\end{tabular}
\caption{Input parameters (mostly from the previous Beta Beam Decay Ring design report~\cite{FP6-FINAL}) above the first line.
  Assumed transversal impedance parameters between the lines. Calculated parameters below the last line. 
  These parameters are the same for the two isotopes, $^{18}$Ne and $^{6}$He.}
\label{tab:drValues}
\end{table}

\vspace{-0.2cm}
\section{Collective Effect Studies}
\label{sec:coll}
\vspace{-0.2cm}


High intensity ion beams, foreseen for the Beta Beam project,  could suffer ``Collective Effects''. 
These are caused by electromagnetic interactions between particles in the beam, 
either with each other directly or through the surrounding environment.
Collective effects could limit the final performance of the accelerators. The studies
of instabilities of all ion beams and all machines in the Beta Beam complex is therefore a crucial 
part of the project. Here we have focused on $^{18}$Ne and $^6$He in the DR.



A particle traveling inside an accelerator leaves electromagnetic fields lagging behind.
Trailing particles feel a force due to the net field caused by all preceding particles. 
The line integral of this force over a certain length (which could be a part of the beam chamber)
gives what the particles see as "wake fields".
If the wake fields last for the duration of the bunch ($\approx$100~ps) particles in the 
``tail'' of the bunch can interact with the wake fields caused by the particles in the ``head'' of the bunch and cause 
{\it single bunch instabilities}. 


The action of the wake fields are described by the {\it wake potential}, $W(t)$, in the time domain and by the 
{\it impedance}, $Z(\omega) = \mathcal{F}\left[W(t)\right]$, in the frequency domain.
This report studies impedances caused by wake fields trapped in cavities of the vacuum chamber,
so called {\it resonance impedances}, $Z^{res}(\omega)$.
If the {\it quality factor} is $Q$~=~$R\sqrt{C/L}$ and the {\it resonance frequency} is $\omega_r$~=~$1/\sqrt{LC}$
the resonance impedance can be modelled as an RLC circuit \cite{Chin:8thJoint} in the transverse plane as
\begin{equation}
Z_{\perp}(\omega) = \frac{R_{\perp}\frac{\omega_r}{\omega}}{1+iQ\left(\frac{\omega_r}{\omega}-\frac{\omega}{\omega_r} \right)}
\label{eq:rlc}
\end{equation} 
where $R_{\perp}$ is the {\it transverse shunt impedance}, assumed to have a value close to RHIC; 2~M$\Omega$/m~\cite{Fischer:2008zzc}.
So far we have only studied short lived resonance wake fields,
i.e. {\it broadband} ($Q$~=~1) impedances in the transverse plane. 
There are many different types of collective effects that could lead to beam instability 
but this study is constrained
to {\it transverse broadband resonance impedances}.
Three different methods have been used to achieve the maximum number of allowed ions per bunch, $N_B^{th}$.

One approach is to use the peak current values of the bunch current and momentum spread as input  
to a coasting beam formula. This gives an expression for the intensity limit that we will call the 
{\it coasting beam equation} \cite{Metral:2004vi} and which for zero chromaticity (as assumed all through this report) is
\begin{equation}
N_{B_{x,y}}^{th} = \frac{32}{3\sqrt{2}\pi}\frac{Q_{x,y}|\eta|\varepsilon_l^{^{2\sigma}}\omega_r}{cZ^2\beta^2R_{\perp}} \ .
\label{eq:cb}
\end{equation}
Here $c$ is the speed of light in vacuum and all other parameters are
given in table~\ref{tab:drValuesIonDep} and \ref{tab:drValues}.

MOSES \cite{Chin:1988xj} solves an integral equation in the frequency domain to give the rise time, $\tau$, of the 
instabilities for different head-tail modes as a function of the bunch intensity. 
The limit, $I_b^{th}$, is given by the most crucial head-tail mode after defining the maximum allowed 
growth rate, $(1/\tau)^{th}$. 
To reach the ion equivalent intensity threshold we divide by a factor $Z$; $\bar{I}_b^{th} = I_b^{th}/Z$. 
The maximum allowed number of ions per bunch is then given by the conversion $N_B^{th}$~=~$T_{rev}\bar{I}_b^{th}/Ze$.

The third method uses the multiparticle tracking code HEADTAIL\cite{headtail} where a bunch of macro-particles is sliced longitudinally and 
the impedance is assumed to be localized at a few positions around the ring. At each impedance location, each slice leaves
a wake field behind and gets a kick by the field generated by the preceding slices. 
The bunch is then transferred to the next impedance location via a transport matrix. 
For the Beta Beam studies the possibility of bunches with  $^{18}$Ne and $^{6}$He  was added to the code. 
An exponential least squared fit 
to the envelope of the vertical oscillation of the mean bunch position
gives the growth rate of the instability. Same as for MOSES the bunch intensity limit, $N_B^{th}$, is reached
when the rise time is shorter than allowed, i.e. $1/\tau > (1/\tau)^{th}$.  

It could be argued that instabilities with the longest rise times should define $N_b^{th}$, i.e. $(1/\tau)^{th} \to 0$. 
However in this report we will take an optimistic approach, assume that slow instabilities can be damped with sextu-
and octupoles and define $(1/\tau)^{th}~=~400~1/s$ for both MOSES and HEADTAIL. 

\begin{figure}[ht]
$\begin{array}{c@{\hspace{0.01in}}c}               
\includegraphics[angle=0, scale= .157]{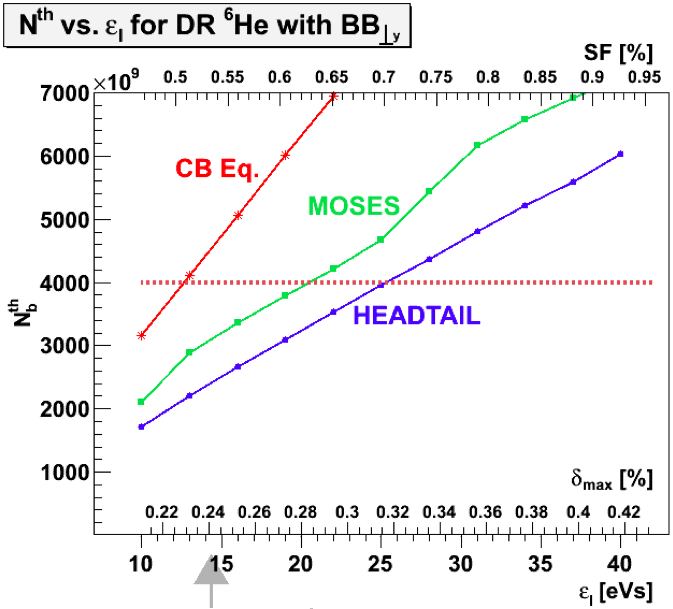}  &  \includegraphics[angle=0, scale= .157]{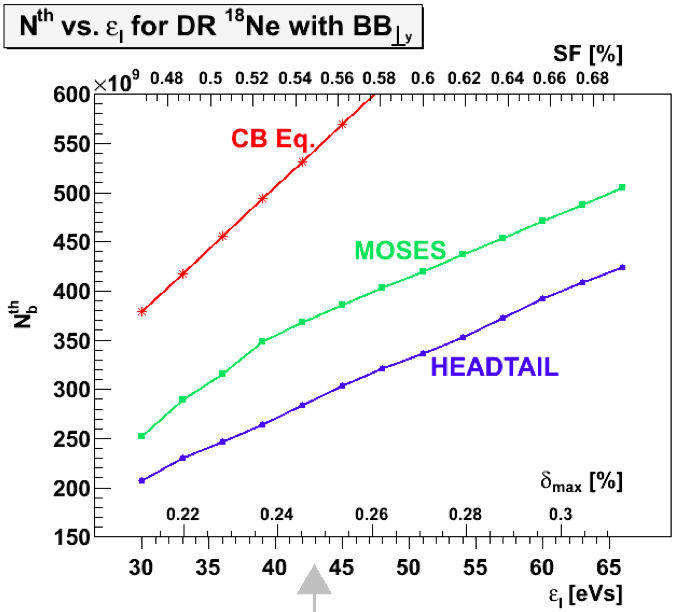}    \\ [-3.5cm]
\multicolumn{1}{l}{\mbox{\bf \ \ \ \ \ \ \ \ \ \ \ \ \ \ \ \ \ \ \ \  \ \ \ \ \ \ \ \  (a)}} &  
\multicolumn{1}{l}{\mbox{\bf \ \ \ \ \ \ \ \ \ \ \ \ \ \ \ \ \ \ \  \ \ \ \ \ \ \ \   (b)}}                      \\ [+3.1cm]           
\includegraphics[angle=0, scale= .157]{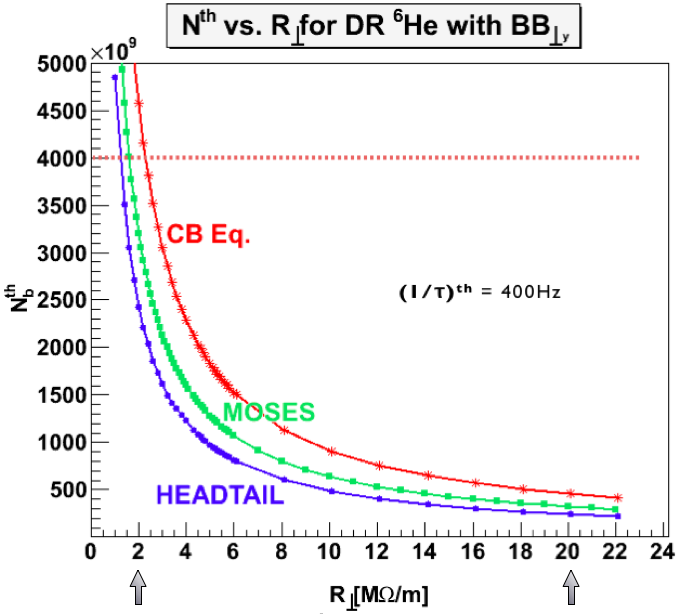}   &  \includegraphics[angle=0, scale= .157]{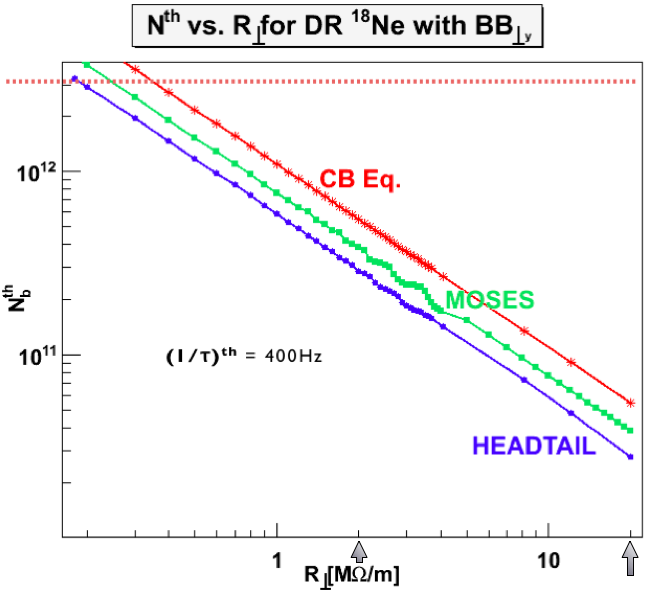}     \\ [-3.6cm]
\multicolumn{1}{l}{\mbox{\bf (c)}}                  &  \multicolumn{1}{l}{\mbox{\bf (d)}}                    \\ [+2.9cm]  
\end{array}$
\caption{$N_B^{th}$ as a function of {\bf (a, b)} the longitudinal emittance and {\bf (c, d)} the transversal shunt impedance
  according to the C.B. eq.~(\ref{eq:cb}), MOSES~\cite{Chin:1988xj} and HEADTAIL~\cite{headtail}. 
  The right (left) column shows the case for the (anti) neutrino emitter, $^6$He ($^{18}$Ne), 
  in the Decay Ring. With the log-log scale in {\bf (d)} we see that $R_{\perp}$~$\approx$~0.2~M$\Omega$/m
  is needed to allow $3\cdot10^{12}$ $^{18}$Ne per bunch.}
\label{fig:scans}
\end{figure}

With the three methods, mentioned above, we studied the effect on the bunch intensity limit, $N_B^{th}$, by changing slightly the longitudinal 
emittance, $\varepsilon_l$, (fig.~\ref{fig:scans} {\bf (a)} and {\bf (b)}) and assuming $R_{\perp}$~=~2~M$\Omega$/m. 
Fig.~\ref{fig:scans} {\bf (a)} shows that according to MOSES (HEADTAIL) increasing $\varepsilon_l$ with about 5~(10)~eVs 
from the working point (indicated by grey arrow) the desired number $^6$He per bunch, $4\cdot10^{12}$, would be acceptable. 
This would however also mean an undesired increase in SF and momentum spread (also indicated in fig.~\ref{fig:scans} {\bf (a)}). 
As can be seen in fig.~\ref{fig:scans} {\bf (b)} the bunch intensity limit for $^{18}$Ne, $3\cdot10^{12}$, is far out of reach when 
$R_{\perp}$~=~2~M$\Omega$/m is assumed. 

Since impedance could improve in modern machines compared to old accelerators a scan over the shunt impedance
was performed to see the impact on $N_B^{th}$
(fig.~\ref{fig:scans} {\bf (c)} and {\bf (d)}). 
Fig.~\ref{fig:scans} {\bf (c)} shows that for a shunt impedance at the level of SPS, $R_{\perp}^{sps}$~=~20~M$\Omega$/m, 
maximum number $^6$He allowed per bunch, according to HEADTAIL and MOSES, is not more than 300$\cdot10^9$. 
For $N_B$~=~$4\cdot10^{12}$ $^6$He per bunch $R_{\perp}$~$\approx$~2~M$\Omega$/m (similar to RHIC) is needed.
However, as can be seen in the log-log scale plot of fig.~\ref{fig:scans} {\bf (d)}  $R_{\perp}$~$\approx$~0.2~M$\Omega$/m 
is needed in the DR to allow $3\cdot10^{12}$ $^{18}$Ne per bunch.


Scans over resonance frequency, $f_r = \omega_r/(2\pi)$, 
and chromaticity, $\xi_y = (dQ_y/Q_y)/(dp/p)$, were also 
performed without any significant relaxation in bunch 
intensity limit, $N_B^{th}$, within realistic ranges of
the scan parameters.


\vspace{-0.3cm}
\section{Conclusions}
\label{conclusions}
\vspace{-0.2cm}

There will be large challenges due to requirements of seemingly insurmountable low transverse broadband impedance of the Beta Beam Decay Ring. 
This study, based on parameters mostly from~\cite{FP6-FINAL} (table~\ref{tab:drValuesIonDep} and \ref{tab:drValues}), 
suggests a reoptimization of the Beta Beam Decay Ring design.   



\vspace{-0.2cm}
\begin{theacknowledgments}
\vspace{-0.2cm}
We are grateful for very useful discussions with and/or working material by
E. Benedetto, A. Chanc\'e, E. Metral, N. Mounet, B. Salvant and E. Wildner. 

We acknowledge the financial support of the European Community under the
European Commission Framework Programme 7 Design Study: EUROnu, Project
Number 212372. The EC is not liable for any use that may be made of the
information contained herein.

\end{theacknowledgments}
\vspace{-0.4cm}

\bibliographystyle{aipproc}   
\bibliography{References}

\end{document}